\documentclass[aps,prl,twocolumn,a4paper,10pt,notitlepage,footinbib,superscriptaddress,showpacs]{revtex4-2}%
\usepackage[english]{babel}
\usepackage[T1]{fontenc}
\usepackage{endnotes}
\usepackage{amssymb,amsmath,amsfonts}
\usepackage{textcomp}
\usepackage{graphicx,color}
\usepackage[dvipsnames]{xcolor}
\usepackage[utf8x]{inputenc}
\usepackage{mathtools}
\usepackage{float}
\DeclarePairedDelimiter{\ceil}{\lceil}{\rceil}

\usepackage{tabularx}

\begin{document}

\title{Competition between local erasure and long-range spreading of a single biochemical mark leads to epigenetic bistability}

\vspace{-0.2 cm}
\author{Marco Ancona}
\affiliation{SUPA, School of Physics and Astronomy, University of Edinburgh, Edinburgh EH9 3FD, United Kingdom}
\author{Davide Michieletto}
\affiliation{SUPA, School of Physics and Astronomy, University of Edinburgh, Edinburgh EH9 3FD, United Kingdom}
\affiliation{MRC Human Genetics Unit, Institute of Genetics and Molecular Medicine, University of Edinburgh, Edinburgh EH4 2XU, United Kingdom}
\affiliation{Centre for Mathematical Biology, and Department of Mathematical Sciences, University of Bath, North Rd, Bath BA2 7AY, United Kingdom}
\author{Davide Marenduzzo}
\affiliation{SUPA, School of Physics and Astronomy, University of Edinburgh, Edinburgh EH9 3FD, United Kingdom}

\date{\today}

\begin{abstract}
\vspace{-0.3 cm}
The mechanism through which cells determine their fate is intimately related to the spreading of certain biochemical (so-called epigenetic) marks along their genome. The mechanisms behind mark spreading and maintenance are not yet fully understood, and current models often assume a long-range infection-like process for the dynamics of marks, due to the polymeric nature of the chromatin fibre which allows looping between distant sites. While these existing models typically consider antagonising marks, here we propose a qualitatively different scenario which analyses the spreading of a single mark. We define a 1D stochastic model in which mark spreading/infection occurs as a long-range process whereas mark erasure/recovery is a local process, with an enhanced rate at boundaries of infected domains. In the limiting case where our model exhibits absorbing states, we find a first-order-like transition separating the marked/infected phase from the unmarked/recovered phase. This suggests that our model, in this limit, belongs to the long-range compact directed percolation universality class. The abrupt nature of the transition is retained in a more biophysically realistic situation when a basal infection/recovery rate is introduced (thereby removing absorbing states). Close to the transition there is a range of bistability where both the marked/infected and unmarked/recovered states are metastable and long lived, which provides a possible avenue for controlling fate decisions in cells. Increasing the basal infection/recovery rate, we find a second transition between a coherent (marked or unmarked) phase, and a mixed, or random, one.
\end{abstract}

\maketitle

\section{I. Introduction}
The field of epigenetics studies how inheritable changes in gene expression can arise without any changes in the DNA (genetics) of a certain organism~\cite{Alberts,Sneppen}. These processes are important, for instance, to explain how different cells of the same eukaryotic organism -- which are genetically identical -- specialise to give rise to different tissues. One well-characterised way to store epigenetic information in a cell is via biochemical modifications, or marks, of the histone octamers which associate with the eukaryotic DNA to form the ``chromatin fibre''~\cite{Cortini}. Existing evidence points to the fact that such modifications are quite dynamic, and are constantly deposited and erased over time by a vast range of epigenetic enzymes, commonly referred to as ``writers'' and ``erasers''.
Focussing on a given genomic region, different parameters can either favour or hinder deposition of a mark along DNA, resulting in an effective ``spreading'' (or erasure) of the mark in that region.
A key fact is also that epigenetic marks are globally lost following cell division, hence they need to be re-established in a robust way at each generation. This is crucial as, for instance, a skin cell needs to retain the epigenetic memory of its state following division (and not to turn into a different type of cell). At the same time, epigenetic plasticity (a cell's ability to generate different epigenetic patterns given suitable cues) is necessary during cellular reprogramming or differentiation from a pluripotent progenitor~\cite{Festuccia2017,Michieletto}.
Single-cell experiments in yeast suggest that stochastic transitions between two epigenetic states (i.e., characterised by the presence or absence of a particular epigenetic mark), can occur either over the cell lifetime, or over a few generations~\cite{Xu,O'Kane}. This is especially the case in ``weakened'' region of the genome, where, for instance, nucleation sites are missing or defected~\cite{Sedighi}. Intermittent switching between two epigenetic states is therefore possible, and is commonly referred to as ``epigenetic bistability''~\cite{Sneppen}.

\begin{figure*}[t]
\includegraphics[width=1.\textwidth]{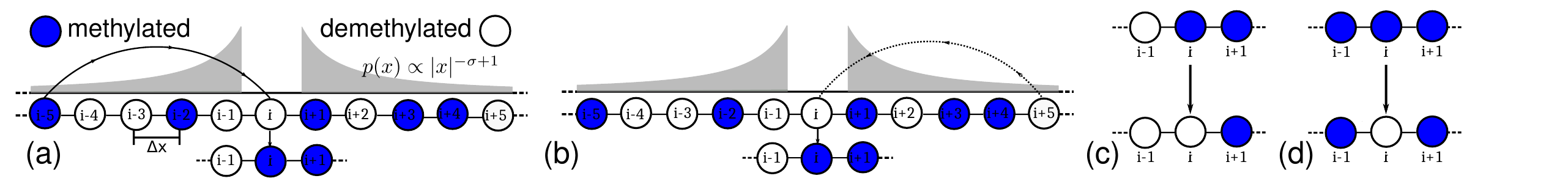}
\caption{\textbf{Microscopic rules.} Representation of the transition rules for methylated/infected sites (blue circles) and demethylated/recovered ones (white circles). \textbf{(a)} Site $i$, which is unmarked in this case, is in long-range contact with site $j$ (here, $j=i-5$). If $j$ is marked, $i$ can become methylated through a long-range ``infection'' at a rate $q_\lambda(1+\lambda)$. \textbf{(b)} If $j$ (here, $j=i+5$) is unmarked, $i$ can still become infected spontaneously, at rate $q_\lambda$. \textbf{(c)} To simulate the erasing process, a marked site $i$ can be recovered to the unmarked state at rate $q_\mu(1+\mu)$ when at least one of the two neighbouring sites is unmarked. \textbf{(d)} A marked site $i$ that is flanked by marked sites can spontaneously loose the mark at rate $q_\mu$.}
\label{fig1}
\end{figure*}

A number of dynamical models have been proposed in the literature to explain the phenomenon of epigenetic memory and bistability~\cite{Cortini,Sneppen,Michieletto,Berry,MichielettoPRX,Jost,Jost2,Adachi,Coli}.
 These previous works have typically considered the interplay between two competing marks: methylation of histone 3 at lysine 9 (H3K9me, or methylation for brevity) which is associated with inactive genes, and acetylation of the same residue (H3K9ac) which is associated with active genes. The kinetic rules used in those models give rise to a competition between the two marks~\cite{Sneppen}, which may in turn yield bistability between a globally active and inactive state~\cite{Micheelsen,Sneppen}, via a symmetry breaking mechanism loosely similar to that through which the Ising model selects its stable state~\cite{Michieletto}. Such systems can then retain memory of their (active or inactive) state even in the presence of an external perturbation -- such as the stochastic loss of a large fraction of the marks during DNA replication~\cite{Sneppen,Michieletto}. We highlight, though, that one-dimensional models exhibiting bistability in this way typically require the addition of one or more intermediate states~\cite{Sneppen,Berry}. Another generic mechanism for bistability, alternative to spontaneous symmetry breaking, is the proximity to a first-order transition in a finite system: in this case, either switching between two states or coexistence can appear near the transition region~\cite{Erdel2016}. Not only can this explain bistability, but, in the context of heterochromatin (inactive chromatin) spreading and inheritance, it gives a natural explanation of the cell fate memory through epigenetic robustness across generations~\cite{MichielettoPRX}. The model we propose here explores this alternative route to bistability. We believe this is relevant to explain bistability and memory even in stretches of chromatin where a single epigenetic mark can be deposited or erased, rather than in ones where multiple marks compete with each other~\cite{Sneppen}.

Our work is structured as follows. In Section II we define the model, listing the steps of our Monte Carlo algorithm (Section IIa). In Section III we show that albeit our model, for suitable parameter choices, can be mapped onto a \textit{Long-Range Directed Percolation} process (LRPD), it also falls, in some other limits, into a totally different ``non-universality'' class, which shares some features with the \textit{Long-Range Voter Model} (LRVM). In Section IV we discuss our findings when the model is studied in a biologically relevant parameter space. It should be noted that, while our model is inspired by the spreading and establishment of epigenetic marks on the genome of cells, our framework is generic and indeed our findings would hold for an epidemic process that follows the same microscopic infection/spreading and recovery/erasure transition rules. The terms ``infected'' and ``recovered'' are frequently used to describe the state of a system in epidemics and in percolation-like models. Since our work shares some features with these models, we will make use of both the terminologies interchangeably throughout the following of this paper. Specifically, we will speak of an infected state or phase to indicate a biochemically marked state or phase, and we will refer to a recovered state or phase to denote an unmarked state or phase.

\section{II. The model}
 
We assume that the spreading of a mark occurs through long-range contacts mediated by the polymer substrate, the chromatin fibre. This polymer -- made of DNA wrapped around histone octamers -- is folded in 3D space and experiments have shown that its contact probability is a power-law distribution~\cite{Aiden,Caselle}:
\begin{equation}
P(l) \propto l^{-(\sigma + 1)}, \qquad l \rightarrow +\infty
\label{levyflight}
\end{equation}
where $l$ is the distance along the chromatin fibre and $ \sigma > 0 $ is the contact exponent. In other words, the spreading of epigenetic marks can be described in terms of a \textit{L\'evy distribution}~\cite{Sneppen}. Interestingly, the contact exponent defining this distribution can depend on cell type, stage of the cell cycle~\cite{Gibcus2018} and on the chromosome considered~\cite{Barbieri}. Therefore, in our model, $\sigma$ will be treated as a ``free'' parameter, and we show that different values yield qualitatively different behaviours. We should mention that, by using the contact probability in Eq.~\eqref{levyflight}, we are assuming that the timescale associated with chromatin relaxation is smaller than, or comparable to, the one related to the spreading of marks. Experiments suggests that, at least in some cases, this is realistic. For instance, in yeast the estabilishment of an epigenetic state can take generations \cite{Xu}, whereas in mice the spreading of heterochromatin occurs at a rate of $\sim 100$ nucleosomes/day \cite{Hathaway}, while the coherent motion of chromatin can be of about $10\mu$m/day \cite{Zidovska} (which is approximately the size of a nucleus).

In order to model the erasure of H3K9me (or methylation) marks we argue that this is more likely to happen at the boundary of an infected domain rather than in the bulk. This difference might arise, for instance, if diffusion of the enzyme responsible for lifting the epigenetic mark is hindered by the strongly crumpled conformation assumed by regions of the genome enriched of such a mark~\cite{Michieletto} or if the enzyme possesses binding sites for both methylated and demethylated histones. Therefore, we describe mark erasure as a ``local'' process and we assume it is more probable in the presence of gradients in the density of methylation marks. 

\subsection{II. A. MC algorithm rules and relevant quantities}
Our model is defined on a lattice of $ N $ sites, which can be either in the demethylated/unmarked/recovered state or the methylated/marked/infected state (the exact terminology used is traditionally dependent on the research field). Each site is associated with a dichotomous variable $m_i $, which can be respectively either $0$ or $1$. The transition rules are specified in the following way (see Fig.~\ref{fig1}):
\begin{equation}
\begin{split}
0_i1_j &\xrightarrow{q_\lambda(1+\lambda)} 1_i1_j \\
0_i0_j &\xrightarrow{q_\lambda} 1_i0_j
\end{split}
\label{write}
\end{equation}

\begin{equation}
\begin{split}
1_i0_{i\pm 1} &\xrightarrow{q_\mu(1+\mu)} 0_i0_{i\pm 1} \\
1_{i-1}1_i1_{i+1} &\xrightarrow{q_\mu} 1_{i-1}0_i1_{i+1} \\
\end{split}
\label{erase}
\end{equation}

In words, if a site $i$ is unmarked ($m_i = 0$), it can become marked at rate $ q_\lambda(1+\lambda)$ when it enters in contact with a methylated site $j$ ($m_j = 1$). Otherwise, if $m_j = 0$, it can convert spontaneously with rate $q_\lambda$. The site $j$ is selected by drawing the distance $|i-j|$ from a normalized power-law distribution:
\begin{equation}
P(|i-j|) = \zeta(\sigma + 1) |i-j|^{-(\sigma+1)} \qquad i \neq j.
\end{equation}
where $\zeta(\alpha) = (\sum_{n=1}^{\infty} n^{-\alpha})^{-1}$. Once the site $j$ is selected, the conversion rate can therefore be written as 
\begin{equation}
q_{0 \rightarrow 1} = q_\lambda(1+\lambda m_j).
\end{equation}

Conversely, if site $i$ is marked ($m_i = 1$), it can become unmarked at rate $ q_\mu (1+\mu)$ if it is at a domain boundary, that is if $m_{i-1}m_{i+1} = 0$. Otherwise, if $m_{i-1}m_{i+1} = 1$ -- i.e., if the site is flanked by two  marked sites -- it can be demethylated with a basal rate $q_\mu$. In other words the sites in the domains bulk are more protected from erasure. The recovery rate can thus be written as
\begin{equation}
q_{1 \rightarrow 0} = q_\mu\left[1+\mu (1-m_{i-1}m_{i+1})\right].
\end{equation}
The algorithm proceeds with random sequential updates, and each timestep (sweep) consists of $N$ conversion trials:

\begin{itemize}
\item[$1)$]{Extract a random integer number between $1$ and $N$, which selects the $i$-th site. If $m_i = 1$, jump directly to step $4$;} 
\item[$2)$]{if $m_i = 0$, choose with probability $1/2$ one of the two directions to attempt a long-range infection event;}
\item[$3)$]{draw a random number $z$ from a uniform distribution between $0$ and $1$, then consider the real number $r = z^{-1/\sigma}$, round it to the smaller positive integer $\ceil r$ so that $j = \mathrm{mod}\{i \pm \ceil r,L\} + \theta(-i \mp \ceil r ) N$, where $\mathrm{mod}\{\cdot,\cdot\}$ indicates the modulo operation between two integers, $ \theta $ is the Heaviside function, and the $\pm$ depends on the direction choosen in step $2$;}
\item[$4)$]{perform a conversion with rate $ q = (1-m_i) q_{0 \rightarrow 1} + m_i q_{1 \rightarrow 0} $.}
\end{itemize}
Note that the values of $ \lambda $ and $ \mu $ are bound by the values $q_{\lambda,\mu}^{-1}-1$, where $q_{\lambda,\mu} \leq 1$. To compare time in simulation units to physical time, one may calculate and compare any of the dimensionless quantities obtained by multiplying time with one of the rates. This means that a time step in our simulation equals the inverse long-range methylation rate (or the inverse demethylation rates at a domain boundary).

\begin{figure}
\includegraphics[width=.5\textwidth]{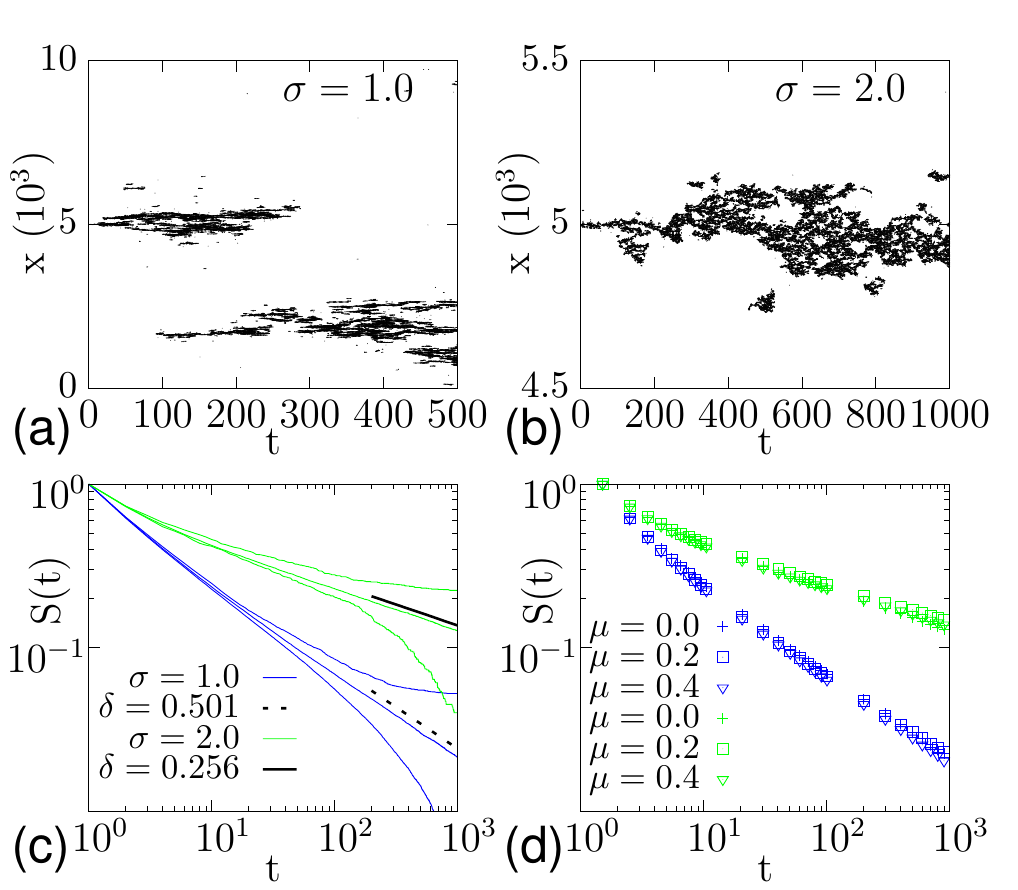}
\caption{\textbf{Kymographs and survival probability for LRDP.} In the first two panels we show two typical kymographs at the transition point, and with $\bar{\mu} = 0.2 $, for respectively \textbf{(a)} $\sigma = 1.0$ \textbf{(b)} $\sigma = 2.0$. Note that there are no very compact domains. \textbf{(c)} Power-law trends of $ S(t) $ for two values of the contact exponent ($\sigma = 1.0$, blue lines, and $\sigma = 2.0$, green lines), with $ \bar{\mu} = 0 $, and the corresponding value of the critical exponent $\delta$ (respectively dashed and solid black lines)}. For clarity, we also show the survival probability just above and below the transition, which display the typical decay and saturation. \textbf{(d)} By varying the value of $ \bar{\mu} $ the survival probability does not change. The critical properties of the system, then, do not depend on the local erasing, but only on the global erasure rate.
\label{fig2}
\end{figure}

In the following, we will mainly consider the rescaled rates $q_\lambda \lambda = \bar{\lambda}$ and $q_\mu \mu = \bar{\mu}$, which, as we will see, play a relevant role in the dynamics of methylation/infection. To describe the global methylation/infection state, we use the order parameter $ m \equiv \langle m \rangle = 1/N \sum_{i=1}^N m_i$, which is the fraction of infected sites in the lattice. Another relevant quantity which is important in the presence of absorbing states is the \textit{survival probability}, $S(t)$, namely the probability that the system has not reached any absorbing states up to time $t$. The survival probability generally depends on the initial condition, and we consider an initial condition with a single infected seed in the following. We will study the behaviour of the survival probability in Section III; since our model entails a non-equilibrium phase transition between the totally recovered ($m = 0$) and infected phases ($m \neq 0$), we pinpoint the location of the transition point by looking for a power-law behaviour of $S(t)$ (see, for instance, Fig.~\ref{fig2}c), that is far more efficient than searching for a discontinuity in the order parameter $m$ or in its derivative.

\section{III. Dynamics with absorbing states}

The first terms in Eqs.~\eqref{write} and \eqref{erase}, $q_\lambda$ and $q_\mu$, are two basal rates which represent respectively spontaneous acquisition and deletion of methylation marks. Within the problem of understanding how epigenetic marks spread, it is natural to set these rates non-zero to account for biological noise and imperfect writing and erasure by the respective enzymes. Yet, it is instructive to consider first the situation in which either one or both the basal rates are zero: these cases lead to interesting physics, and help to better understand the behaviour of the general system.

If any of the basal rates is zero, there are absorbing states in the dynamics of the system. If $q_\lambda \rightarrow 0 $  there is one absorbing state ($m=0$); if $q_\lambda,q_\mu \rightarrow 0 $, the absorbing states are two (either $m=0$, or $m=1$). As the absorbing states are configurations from which it is not possible to escape, detailed balance is violated and the resulting system is out-of-equilibrium.

\begin{figure*}
\includegraphics[width=1.\textwidth]{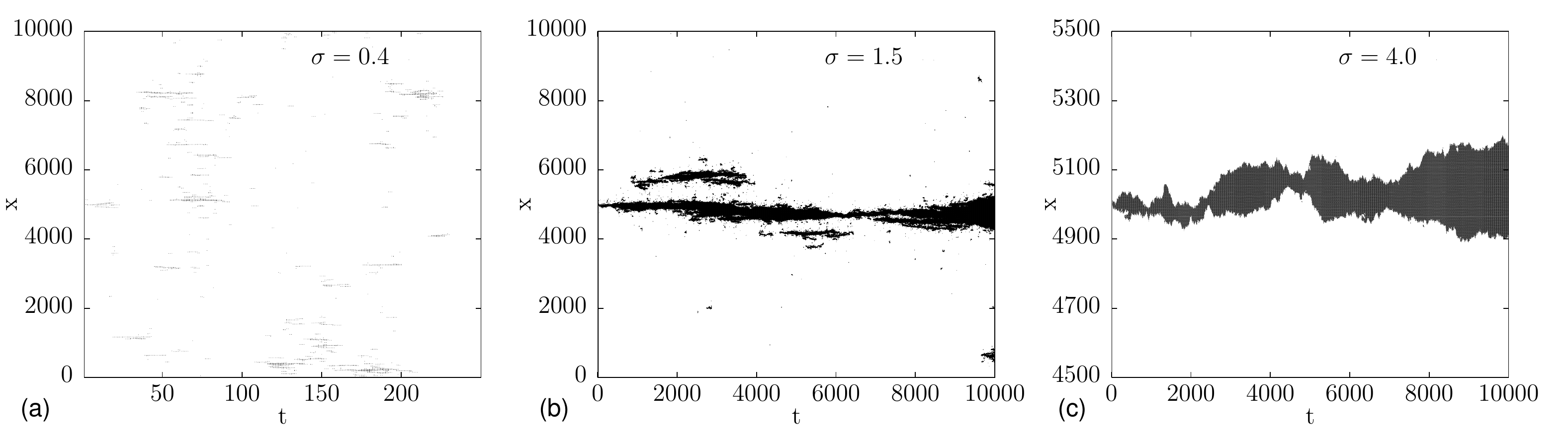}
\caption{\textbf{Kymographs at the transition point.} Representative kymograph of the infection/methylation profile, when the system is at the critical point ($\kappa = 0$ in Eq.~\eqref{LRCDP}, $S(t)$ is a power-law). \textbf{(a)} For $ \sigma = 0.4 $ no compact domains can live for long times, and the systems falls into the absorbing states after few hundreds timesteps. \textbf{(b)} For $ \sigma = 1.5 $ the main domain remains compact and lives for the entire duration of the simulation. Other domains can be created and erased. \textbf{(c)} For $ \sigma = 4.0 $ the only active compact domain is generated by one infected seed. it can live for a virtually infinite time.}
\label{fig3}
\end{figure*}

As we shall show, our model displays non-equilibrium phase transitions of different type, depending on the number of the absorbing states (either one or two), which reflects the symmetries involved in the corresponding effective action (see Appendices). This model is respectively either in the LRDP non-universality class -- and characterised by a second-order phase transition with a $\sigma$-dependent exponent -- or, alternatively, can be mapped onto the \textit{Long-Range Compact Directed Percolation} (LRCDP),  -- characterised by a first-order phase transition with a discontinuity in the average infected/methylated fraction of sites $m$.  It is well known that the standard Compact Directed Percolation (CDP) -- i.e., the case with short-range interactions -- falls in the same universality class of the \textit{Standard Voter Model} (SVM) in $ d = 1 $. Moreover, critical exponents can be exactly computed, since the dynamics can be seen as an annihilation problem between two random walkers~\cite{Peliti}.
 
There are also studies of the annihilation of L\'evy flights~\cite{Vernon,Vernon2} and of the Voter Model with long-range interactions/infection~\cite{Albano}: however, their aim is distinct from the one we pursue here. In particular, these works direct their efforts towards the computation of the dynamical exponent $ \alpha $ characterising the decay of the density of infected sites, $\rho$, with time, i.e., $ \rho \sim t^{-\alpha} $. This requires simulations which start with uniform density or fully disordered configurations. Here, instead, we focus on the survival probability exponent $\delta(\sigma)$ in simulations where the initial configuration has a single infected/methylated site (single seed simulations). Such condition is typically used to compute the dynamical exponent $ \delta $, since one expects a power law dependence of the survival probability with time -- i.e., $ S(t) \sim t^{-\delta}$ -- for $ t \gg 1 $. Our aim is to: (i) check that in the limit of $q_\lambda \rightarrow 0 $ ($q_\mu $ finite) the contact process belongs to the LRDP non-universality class (Section IIIA), and (ii) show that in the limit $q_\lambda, q_\mu \rightarrow 0 $ our model falls in a different non-universality class, which is the LRCDP one (Section IIIB). To the best of our knowledge, the latter case has never been studied systematically, and for this case we compute the critical exponent $\delta$.

\subsection{III. A. Limit $ \mathbf{q_\lambda \rightarrow 0}$}

In the limit where $ q_\lambda \rightarrow 0$ and $ \lambda \rightarrow +\infty$ such that the product $ q_\lambda\lambda \equiv \bar{\lambda}$ is finite and $ q_\mu > 0$, the system cannot recover from the totally demethylated state. The \textit{unique} absorbing state is defined by $m = 0$.  In this limit, our epigenetic model can be recast as an infection model with long-range infection and a single absorbing state. As such, we expect it to be in the same universality class of long-range DP. The phenomenological equation for LRDP reads~\cite{Howard}:
\begin{equation}
\partial_t m = \kappa m - g m^2  + D_A \nabla^{\sigma} m + \sqrt{\Gamma m}\zeta
\label{LRDP}
\end{equation}
where, in general, $ \kappa = \kappa(\bar{\lambda},\bar{\mu},q_\mu,\sigma) $, $D_A$ is the coefficient of anomalous diffusion, $\nabla^{\sigma}$ denotes a fractional derivative, and $ \Gamma $ is the strength of the noise $ \zeta $. This equation predicts two different stable states in the mean field treatment: for $ \kappa < 0 $ the stable state is the absorbing one, defined by $ m = 0 $, whereas for $ \kappa > 0 $, the stable state is $ m = \kappa/g $. Clearly the system is critical at $ \kappa = 0 $. 

In agreement with our expectation, the exponents found for simulations $ 10^3 $ timesteps long on a $ N = 10000 $ lattice give results in accordance with those found for LRDP in Ref.~\cite{Howard} (Fig.~\ref{fig2} in this paper should be compared with Fig.~2 and Fig.~4 in Ref.~\cite{Howard}). Two representative kymographs are shown in Fig.~\ref{fig2}, for $\sigma = 1.0,2.0 $ at the critical point. Note that, when the infection process is long-range, multiple infected nuclei can be generated very far away from the original seed (see Fig.~\ref{fig2}a). Conversely, when it is short-range, the infected domains are more gathered around the original seed position (see Fig.~\ref{fig2}b). In any case, domains are not compact -- i.e. there are holes within every single domain. 

In Fig.~\ref{fig2}c the power law behaviour of $S(t)$ at criticality is shown, and we find $ \delta(\sigma = 1.0) \simeq 0.501 $ and $ \delta(\sigma = 2.0) \simeq 0.256 $, which are compatible with the exponents found in \cite{Howard} . We cannot compute the survival probability for $ \sigma \leq 0.5 $, that is the mean-field regime (see Appendix B), since the long-range correlation are too important, and the finite-size effects are not negligible. Finally, we show that the exponent is independent of the value of $ \bar{\mu} $ (see Fig.~\ref{fig2}d), which just modifies the value of $q_\mu$ at which the system is critical. In other words, models with only gradient-dependent (or ``local'') recovery are in the same universality class as models with global recovery terms at a fixed $\sigma$. This is consistent with the relevant terms in the Reggeon effective field theory (see Eq.~\eqref{mod_actionLRDP}), which are the same for every value of $ \bar{\mu} $ since the structure of the absorbing state is given, as shown in Appendix B. 

\subsection{III. B. Limit $ \mathbf{q_\lambda,q_\mu \rightarrow 0}$}

In the limit in which $q_\lambda, q_\mu \rightarrow 0$, and $ \lambda, \mu \rightarrow +\infty $, in such a way that the quantities $ q_\lambda\lambda, q_\mu\mu $ remain finite, the second reactions in Eqs.~\eqref{write} and \eqref{erase} are both suppressed, and the methylation/demethylation dynamics takes place only if the system is not fully demethylated/methylated. Indeed, in this case there are \textit{two} absorbing states defined by $m = 0,1$. The presence of two rather than one absorbing states modifies the universality class, so that the system is no longer equivalent to LRDP. Again, if we define $ q_\lambda\lambda = \bar{\lambda}$ and $q_\mu\mu = \bar{\mu}$, the prototypical phenomenological equation for the field $m(x,t)$ is:
\begin{equation}
\partial_t m = \kappa m(1-m) + D_A \nabla^{\sigma} m + \sqrt{\Gamma m(1-m)}\zeta \, ,
\label{LRCDP}
\end{equation}
where $\kappa$ is a generic function of the microscopical rates $\bar{\lambda}$ and $\bar{\mu}$. Unlike Eq.~\eqref{LRDP}, the effective action built from Eq.~\eqref{LRCDP} has another symmetry under the transformations $ m \rightarrow 1-m $ and $\hat{m} \rightarrow -\hat{m}$ (see Appendix A), which is the reason why this variant of the model behaves differently from LRDP. Eq.~\eqref{LRCDP} predicts that for $ \kappa < 0 $ the completely demethylated state $ m = 0$ is stable, whereas for $ \kappa > 0 $ the stable state is $ m = 1 $. At $ \kappa = 0 $ the system undergoes a discontinuous transition for every value of $ \sigma $. \\

In Fig.~\ref{fig3} are shown three representative kymographs of the infection dynamics at the transition, for different values of the contact exponent $\sigma$. These kymographs are obtained by evolving our microscopic model, detailed in Section II, choosing $\bar{\mu}/\bar{\lambda}$ so that the system is at the transition between fully marked and unmarked phases. For $ \sigma = 0.4 $ we see that no compact domains can form; instead, the infected/methylated domains are full of holes due to the long-range infection process, and at late times the system falls into the absorbing state $ m = 0 $ (Fig.~\ref{fig3}a). 
For $ \sigma = 1.5 $ a compact domains opens in the middle of the lattice: this can also create additional domains by infecting regions of the lattice far away from it, since the interactions are still long-range (see Fig.~\ref{fig3}b). 
For $ \sigma = 4.0 $ the infection profile looks very similar to a typical CDP kymograph, with a single fluctuating compact domain which can virtually persist forever (Fig.~\ref{fig3}c).  

\begin{figure*}
\includegraphics[width=1.\textwidth]{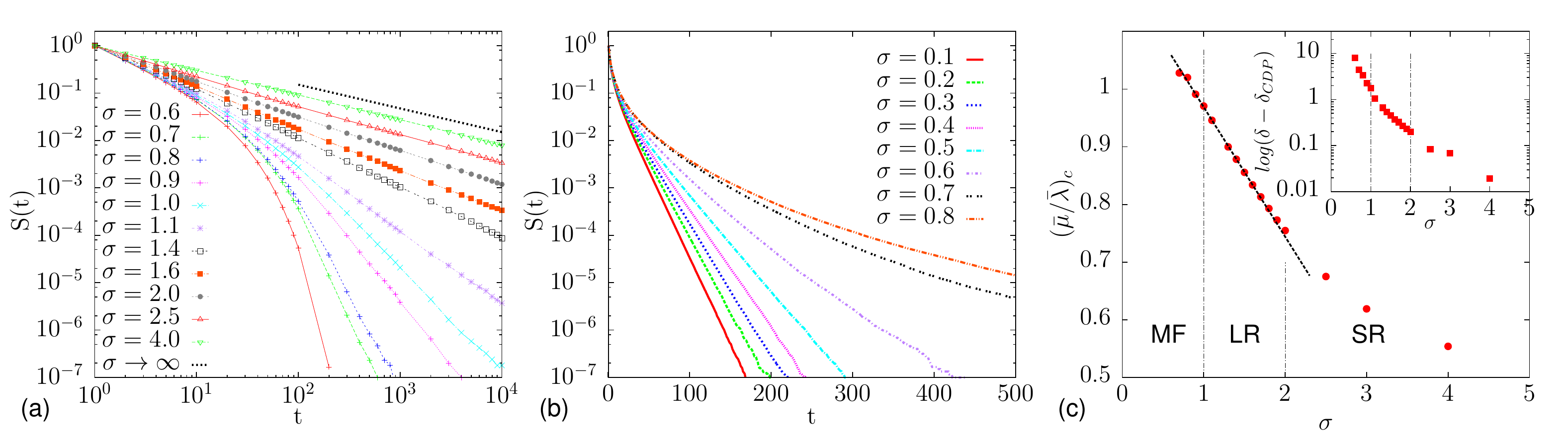}
\caption{\textbf{Survival probability and transition point.} \textbf{(a)} In this panel we show the survival probability in a log-log plot, varying $ \sigma $ in the range $ \left[0.6,4.0\right]$. For $ \sigma > 2 $ the exponent tends to the value predicted by the CDP model ($\delta = 0.5$). As $ \sigma $ decreases the exponent $ \delta $ increases, and the transient is longer. \textbf{(b)} For $ \sigma < 0.6 $ no power laws are observed. Instead, the decay is exponential (here we plot the survival probability in a log-linear scale). For $ \sigma \gtrsim 0.6 $ the exponential decays disappears, instead a power law behaviour takes place. \textbf{(c)} The location of the transition point increases as $ \sigma $ decreases. For $ \sigma > 2 $ (short-range regime, SR) the value of $(\bar{\mu}/\bar{\lambda})_c$ decays slowly to $ 0.5 $. For $ 1 < \sigma < 2 $ (long-range regime, LR) $(\bar{\mu}/\bar{\lambda})_c$ the transition point scales linearly with the contact exponent. For $ \sigma < 1 $ (mean-field regime, MF) the location of the transition point is still compatible with the linear scaling, but cannot be computed for $ \sigma < 0.6 $. In the inset we show the divergence of the survival exponent $ \delta $ as a function of $ \sigma $.}
\label{fig4}
\end{figure*}

In Fig.~\ref{fig4}a we show the survival probability as a function of time, for different values of $ \sigma \geq 0.6 $. For $ \sigma > 2 $ the slope approaches the value predicted for the SVM ($\delta = 0.5 $). Indeed, one would expect that for $ \sigma \geq 2 $ the substitution $ \nabla^\sigma \rightarrow \nabla^2 $ holds and Eq.~\eqref{LRCDP} becomes the well-known equation of CDP. This is, however, not the case: the system displays a smooth crossover between the long range and the short range behaviour, with the latter being re-established only at $ \sigma \gtrsim 4.0$, as shown in Fig.~\ref{fig4}c and its inset. A similar crossover has been found in other models with L\'evy flights, such as the LRDP~\cite{Howard} and the LRVM~\cite{Vernon}. Interestingly, we note that approximately the same value for the crossover which we obtain was found in Ref.~\cite{Albano} for the LRVM. Moreover, it is known that the SVM and CDP belong to the same universality class in $ d = 1 $, but not in other physical dimensions. 

For smaller values of the contact exponent than those considered in Fig.~\ref{fig4}a (i.e., for $ \sigma < 0.6 $) no power law can be detected in our simulations, see Fig.~\ref{fig4}b. This fact may be due to finite-size effects, which can be strong for such long interactions. Nevertheless, we observe that, after a very small transient, the survival probability decays exponentially, $ S(t) \sim \mathrm{exp}(-k(\sigma)t) $. This exponential decay is absent for $ \sigma $ greater than $ 0.6-0.7 $. Such crossover is highlighted in Fig.~\ref{fig4}b, where we replot the survival probabilities for $ \sigma = 0.7$, $0.8$ in log-linear scale, which do not present any exponential decay. This change in behaviour is compatible with an analysis of the Reggeon field theory (see Appendix A), which predicts a mean-field behaviour for $ \sigma < 1 $, as the real dimension $ d $ is above the upper critical dimension $ d_c = \sigma $.
  
We close this Section by discussing how we can map the microscopic rates $ \bar{\lambda} $ and $ \bar{\mu} $ to the effective parameter $\kappa$ entering the phenomenological theory in Eq.~\eqref{LRCDP}. To do so, we note that, in the limit $ \sigma \rightarrow +\infty $, the only sites which can be modified are those at the boundary of the domain. Therefore $ \kappa = 0 $ only if $ \bar{\mu} = \bar{\lambda}/2 $, and we conjecture the following form for $ \kappa(\sigma) $:
\begin{equation}
\kappa(\sigma) = \left(1 - \frac{\bar{\mu}}{\bar{\lambda}} f(\sigma)\right)
\label{rate}
\end{equation}
where $ f(\sigma) $ is a monotonic function which satisfies $ f(+\infty) = 1/2 $. Such behavior is validated by simulations: in Fig.~\ref{fig4}c we show the location of the transition point as a function of $ \sigma $. For $ \sigma < 2 $ the transition point scales linearly with the contact exponent, whilst for $ \sigma > 2 $ it decays slowly towards $  \bar{\mu}/\bar{\lambda} = 0.5 $. This indicates that $ \sigma = 2 $ is the crossing point between the short-range regime and the long-range behaviour. Interestingly, the value of $ \delta $ seems to diverge for smaller and smaller values of $ \sigma $ (see inset in Fig.~\ref{fig4}c); such divergence signals the crossover to the exponential decay regime shown in Fig.~\ref{fig4}b.

\section{IV. Phase diagram of the system, and connection with epigenetics}

\begin{figure*}
\includegraphics[width=1.\textwidth]{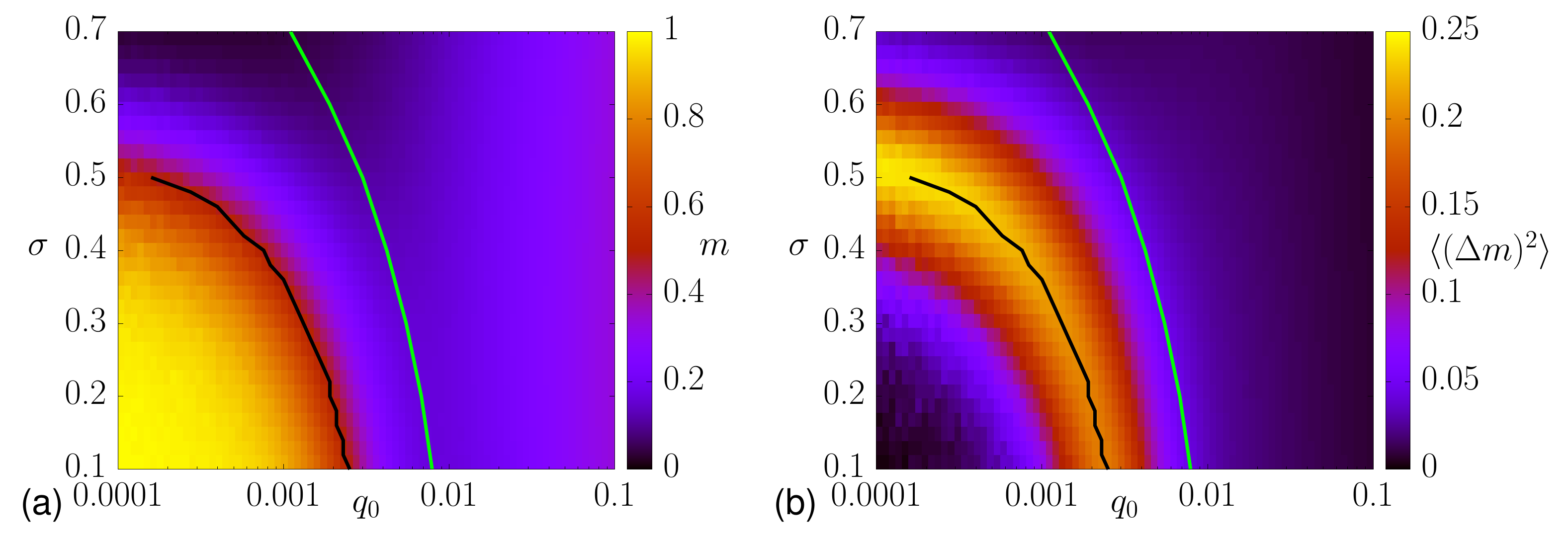}
\caption{\textbf{Phase diagram} \textbf{(a)} Phase diagram as a function of $q_0$ and $\sigma$ for a system with $N=100$. A transition between a marked and an unmarked regime can be seen; the black line correspond to systems where $\langle m\rangle=0.5$. The green line instead separates a coherently unmarked regime from the disordered, or mixed, regime.  \textbf{(b)} Variance of the fraction of marked sites in the system, $ \langle (\Delta m)^2 \rangle$. This quantity peaks close to the transition between the marked and unmarked regimes highlighting a region of enhanced bistability. }
\label{fig5}
\end{figure*}

We now consider a more general case where all parameters are non-zero, so that there are no absorbing states in the system ($q_\mu,q_\lambda\ne 0$). This is likely to be more realistic as it includes the case where there is some generic biological noise. In line with previous models for bistability in epigenetic patterns~\cite{Sneppen,Micheelsen}, we consider a small chromatin region, with $N=100$ beads/nucleosomes. This is a realistic size to study, for instance, stochasticity in epigenetic domains in yeast~\cite{Sneppen,Erdel2016}, or a single chromatin domain of about $20000$ base pairs in larger genomes (as in mammals). 

To render the exploration of parameter space feasible, we set $ q_\mu = q_\lambda = q_0 $, and $ \bar{\lambda} = \bar{\mu} = 1-q_0 $, and we study the behaviour of the model by varying the parameters $ q_0 $ and $ \sigma $. Physically, $q_0$ can be viewed as a temperature-like parameter that regulates the baseline marking/unmarking rates.
Additionally, the biologically relevant values of $\sigma$ are between $0$ and $1$ \cite{Shinkai}. As this exponent should be associated with the looping probability of a polymer representing the chromatin fibre~\cite{BrackleyPRLNonEquilibriumChromosomeLooping}, the value of $\sigma=0.5$ corresponds to looping of a random walk, $\sigma \simeq 1$ corresponds to looping of a self-avoiding walk~\cite{BrackleyPRLNonEquilibriumChromosomeLooping}, whereas $\sigma\simeq 0$ describes the decay of contact probability with genomic distance in a crumpled (or fractal) globule~\cite{GrosbergCrumpledGlobule,MirnyCrumpledGlobule}. 

In Fig.~\ref{fig5}a we present the phase diagram obtained numerically for $ N = 100 $, for different values of $q_0$ and for $\sigma$ between $0.1$ and $0.7$. To find this phase diagram, we used a truncated L\'evy distribution to simulate long-range infection -- i.e. $P(|i-j|) = A|i-j|^{-(\sigma+1)}$, where $ A = \sigma/(1-(N/2)^{-\sigma}) $. This procedure was employed in order to limit boundary effects and it can be shown that it simply shifts the location of the transition line to a slightly smaller value of $\sigma$. 

The phase diagram in Fig.~\ref{fig5}a shows three distinct phases. First, for small value of $q_0$, although no real absorbing state is present, the system still tends to reach a typical state with large $\langle m\rangle$ (methylated regime), or with small $\langle m\rangle$ (demethylated regime). For sufficiently small values of $\sigma$, or equivalently sufficiently long interaction range, long-range methylation dominates over demethylation (bottom left region in the phase diagram). For larger $\sigma$, the fact that the erasure is more likely at domain boundaries tilts the balance in favour of demethylation, and $\langle m \rangle \simeq 0$ (top left region in the phase diagram). Finally, at sufficiently large values of $q_0$, mean field theory applies. The latter predicts that methylation and demethylation should balance giving $m=1/2$ in steady state (as we consider $\bar{\lambda}=\bar{\mu}$). This is the mixed regime, where methylated and demethylated sites coexist in a disordered system, and it is found to the right of the green line in Fig.~\ref{fig5}a. Note that, by using a truncated L\'evy distribution, one can also extend the phase diagram in Fig.~\ref{fig5} to negative value of $\sigma$. However, for $N=100$, we do not expect any different qualitative behaviour from the diagram line $\sigma = 0.1 $, since we are in the mean-field limit for which all the sites connect with each other. Moreover, such values for $ \sigma$ would be forbidden in the limit $N \to \infty$, since Eq.~\eqref{levyflight} would not be integrable.

\begin{figure}
\includegraphics[width=.5\textwidth]{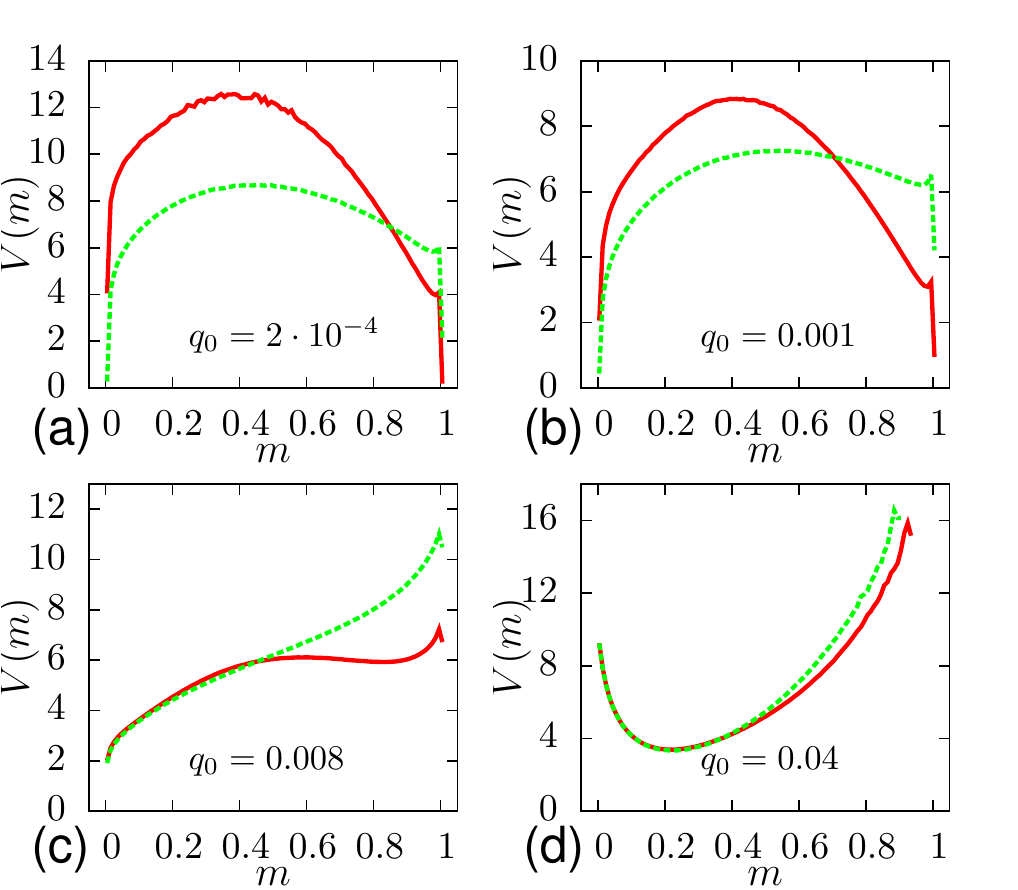}
\caption{\textbf{Methylation effective potentials.} In this figure we show some representative plots of the effective potential $V(m)$, for $ \sigma = 0.2 $ (red solid line) and $ \sigma = 0.6 $ (green dashed line). For $\sigma=0.2$, the increasing values of $q_0$ we consider cross both the marked/unmarked transition line and the coherence transition line. For $\sigma=0.6$, the values of $q_0$ we consider cross only the coherence transition line. \textbf{(a)} For small $ q_0 $ the distribution is bistable, with the global minimum located at $m = 1$ and $ m = 0 $, respectively for $ \sigma = 0.2 $ and $ \sigma = 0.6 $. \textbf{(b)} The increase of $ q_0 $ produces a change in the weights associated with the two minima. \textbf{(c)} After the transition point the global minimum becomes $ m = 0$ in both the cases even if for $ \sigma = 0.6 $ we are already beyond the coherence transition line. \textbf{(d)} For $ q_0 = 0.04$ the system is in the mixed phase and the resulting unimodal distributions depend only very weakly on $ \sigma $, since~\eqref{levyflight} would not be normalized.}
\label{fig6}
\end{figure}

The line separating the coherent -- either marked or unmarked -- regime from the mixed one (which we call the coherence transition line) can be mapped out by analysing the probability distributions of $m$ in steady state, for different parameter values, as shown in Fig.~\ref{fig6}. In particular, in Fig.~\ref{fig6} we show the effective potentials $ V = -\mathrm{log}(P(m)) $-- where $P(m)$ is the probability distribution function for $m$ in steady state -- in the different regions, for the representative values of $ \sigma = 0.2$ and $0.6$. 
In the methylated phase/regime (for $\sigma = 0.2 $, $ m \sim 1 $) the global minimum is located at $ m = 1 $ (see Fig.~\ref{fig6}a). For $ \sigma = 0.6 $, we are in the demethylated phase, and the global minimum is now located at $ m = 0 $. Increasing $q_0$ at small $\sigma$, the system first becomes demethylated (Fig.~\ref{fig6}c, $\langle m \rangle \sim 0.3 $), while for sufficiently large values of $ q_0 $ the effective potential has a global minimum for $0<m<1$ (Fig.~\ref{fig6}d), and the system is in the mixed phase. The green line in Fig.~\ref{fig5}a provides the boundary between the coherent regime (minimum at either $m=0$ or $m=1$) and this mixed phase/regime. It should be noted that, outside the mixed regime, we always find two local minima at $m=0$ and $m=1$, so that both these two states are always at least metastable in the coherent regime. Besides this widespread bimodality, there is a robust bistability region close to the transition between the methylated and the demethylated regime: this is apparent from Fig.~\ref{fig5}b, which shows the variance $\langle (\Delta m)^2 \rangle$ as a function of $q_0$ and $\sigma$. Bistability arises due to the proximity to the methylated-demethylated transition. Because the $P(m)$ distributions are always bimodal, the transition is sharp and first-order like, hence coexistence (and bistability) naturally arise near the critical line. In the bistable region, therefore, our model predicts that epigenetic domains should be highly stochastic and may switch over time. 

To better understand our simulation results, we now discuss a simple analytically tractable approximation which is inspired by the stochastic field theory for long-range compact directed percolation, Eq.~\eqref{LRCDP}. Neglecting the spatial dependency of the order parameter (gradient term), we can write down a dynamical equation for the global methylation in our model, $m$, as follows,
\begin{equation}
\partial_t m = q_0 (1-2m) + \kappa m(1-m) + \sqrt{Q m(1-m)}\zeta.
\label{eqModel}
\end{equation}
The first term in Eq.~\eqref{eqModel} allows \textit{recovery} from the $m=0$ and $m=1$ states, which are then no longer absorbing state. Additionally, $ Q $ is the strength of the multiplicative noise, and $ \zeta $ is a Gaussian random variable, with $ \langle \zeta(t) \rangle = 0$ and $ \langle \zeta(t)\zeta(t') \rangle = \delta (t-t') $. Note that we have neglected also the basal (additive) noise. We want to solve the Fokker-Planck equation associated with Eq.~\eqref{eqModel}, for the probability $P(m,t)$ that the system has a global methylation $m$ at time $t$. This equation reads as follows

\begin{equation}
    \begin{split}
    \partial_t P(m,t) &= \partial^2_{m} \left[ \frac{Qm(1-m)}{2} P(m,t) \right] \\
	& - \partial_m \left[(q_0 (1-2m)P(m,t)\right] \\ 
	& - \partial_m \left[\kappa m(1-m))P(m,t)\right] \\
	& \equiv -\partial_m J(m). 
    \end{split}
	\label{FPsol}
\end{equation}

\begin{figure}
\includegraphics[width=.5\textwidth]{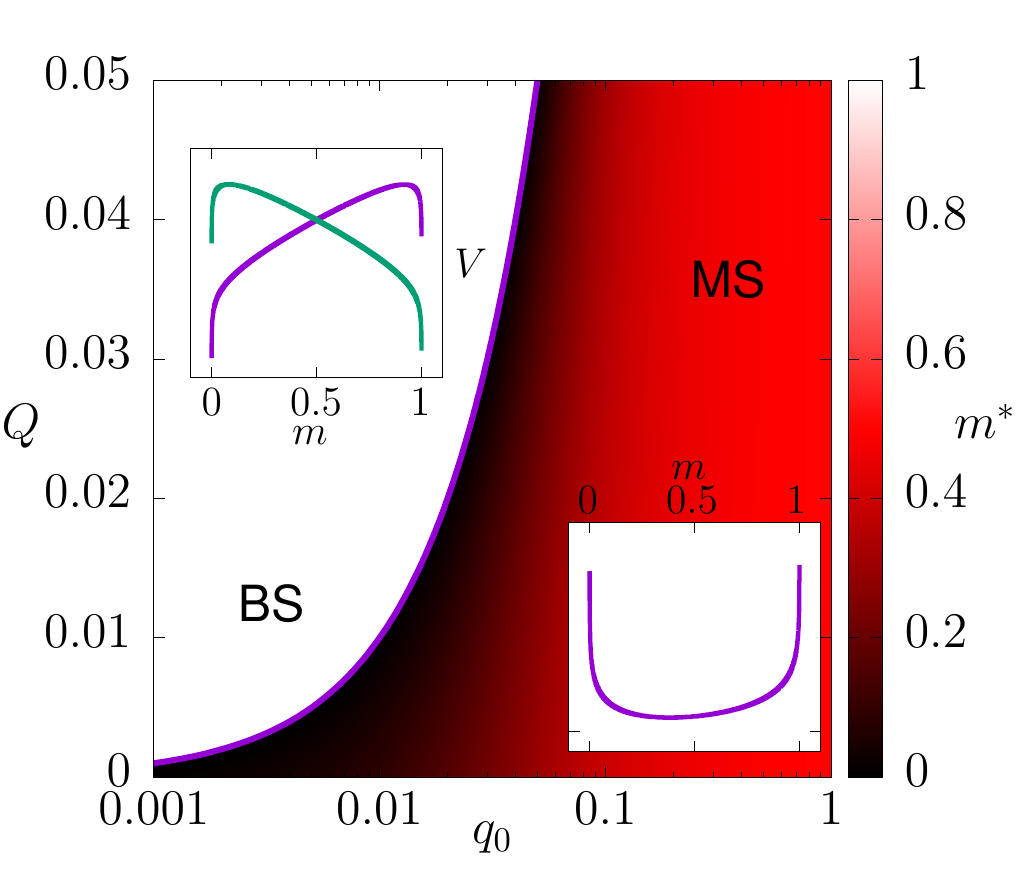}
\caption{\textbf{The bistable-unimodal transition, or the coherence transition line.} This phase diagram displays the transition between a bistable (BS), or coherent, phase and a monostable (MS), or mixed phase, as they are predicted by the mean-field theory. In the first phase (white) the two minima of the effective potential $ V(m)$ are $ m = 0,1 $. In inset we show a typical bistable potential, for $q_0 = 0.001$, $Q = 0.01$, in both the cases $\tilde{\kappa} > 0$, green line, and $\tilde{\kappa} < 0$, purple line ($|\tilde{\kappa}| = 0.1$). In the latter phase only one minimum is present, and its location $ m^*$ varies from $ 0 $ close to the transition to $ 0.5 $, for $ q_0 = 0 $ (if $\tilde{\kappa}$ is negative). In inset we show a typical unimodal potential, for $q_0 = 0.1$, $Q = 0.01$ and $ \tilde{\kappa} = -0.1 $. Colors in the unimodal (mixed) regime are related to the value of $ m^* $ for a fixed value of $ \tilde{\kappa} = -0.1 $. }
\label{fig7}
\end{figure}

By imposing no-flux boundary condition, $ J(m) = 0$, we obtain the following equation for the steady state distribution $P(m)$:

\begin{equation}
    \begin{split}
   \partial_m K(m) = \frac{2\left[q_0 (1-2m) + \kappa m(1-m)\right]}{Q\left[m(1-m)\right]} K(m)
    \end{split}
	\label{FPsol}
\end{equation}
where we set $P(m) \equiv (Q/2)m(1-m)K(m)$.
Then, the stationary probability distribution reads:

\begin{equation}
    \begin{split}
    P(m) &\sim \frac{1}{m(1-m)} \times \\
&\mathrm{exp} \left( 2\int^m \frac{q_0(1-2m') + \kappa m'(1-m')}{Qm'(1-m')} dm' \right)
    \end{split}
	\label{FPsol}
\end{equation}
After some algebra, we find that the system is therefore described by the following effective potential $ V(m) $:
\begin{equation}
    V(m) \propto (1-q_0){\tilde{\kappa}} m - (q_0-Q)\mathrm{log}\left[m(1-m)\right]
	\label{FPpot}
\end{equation}
Since $ \bar{\lambda} = \bar{\mu} = 1-q_0 $ in our microscopic model, we have assumed $ \kappa = (1-q_0)\tilde{\kappa} $. Therefore, there are three parameters, which are fundamental in the mean-field treatment, namely $q_0$, $ \kappa $ and $Q$. For $ q_0 < Q $ $P(m)$ is bimodal, since $ V(m) $ is minimised for $ m = 0,1 $ -- where $ V(m) $ diverges. Conversely, for $ q_0 > Q $, $ V(m) $ becomes unimodal and it is minimised for 
\begin{equation}
    m^* = \frac{1}{2}-\frac{x}{\tilde{\kappa}}+\frac{|\tilde{\kappa}|}{\tilde{\kappa}} \sqrt{\frac{1}{4}-\left(\frac{x}{\tilde{\kappa}}\right)^2}
	\label{FPpot}
\end{equation}
where $ x \equiv (q_0-Q)/(1-q_0) $. Thus, the line $ q_0 = Q $ determines a transition line separating the region where the probability is bimodal, and the system is coherent (here with $m=0$), from the unimodal region where there is a single stable state. Note that $\tilde{\kappa}$ determines the methylation level in the latter case: if $\tilde{\kappa} > 0$ then $m^* > 1/2$, otherwise $ m^* < 1/2$. In all cases, $m^*$ approaches $ 1/2 $ as $ q_0 $ tends to $1$. In Fig.~\ref{fig7} we present a sketch of the line separating the bimodal (epigenetically coherent) from the unimodal/mixed regime according to the analytical theory in Eq.~(\ref{eqModel}). Note that such boundary is independent of the value of $ \tilde{\kappa} $. 
Additionally, if we postulate that the sign of the linear term in Eq.~(\ref{FPpot}) changes, for instance with $\sigma$, $\bar\lambda$, and $\bar{\mu}$, we can describe cases where the coherent regime may correspond to either a methylated or demethylated phase (top-left inset in Fig.~\ref{fig7}).

\section{V. Discussions and conclusions}

In summary, we have proposed here a simple 1-dimensional stochastic model for the dynamics of mark spreading and erasure inspired by the problem of how the epigenetic marks spread along the genome. This is distinct from previous models~\cite{Sneppen,Micheelsen,Jost,Berry}, which typically consider the case where two or more different marks compete within the same region. The case studied here is more appropriate for the study of stochastic gene silencing in yeast, where methylation of H3K9 by, e.g. Clr4, is counteracted by an active erasure process by, e.g. Epe1~\cite{Allshire}. Within our model, methylation is viewed as a long-range ``infection'' process mediated by the looping of the chromatin substrate. Importantly, while infection is long range, we assume that the erasure/recovery occurs locally and is enhanced at domain boundaries. This assumption is motivated, for instance, by the idea that a highly folded or crumpled heterochromatin domain may be difficult to access by a demethylase protein. On average, the most accessible sites of such a domain are thus the ones close to the domain one-dimensional ends. Our local recovery rule may also be enacted by a protein that has binding sites for both marked and unmarked sites and thus is more frequently recruited at sites with gradient in the density of marks.

In our microscopic model, we varied the exponent $\sigma$, regulating the global folding of the chromatin polymer (and hence the range of the infection process), as well as the baseline rate of spontaneous methylation/demethylation, $q_0$. For simplicity, we have taken the methylation and demethylation rates to be equal, and for efficiency we have assumed the rate of long-range methylation and of demethylation at a domain boundary to both equal unity. The phase diagram in the $(q_0,\sigma)$ plane shows two distinct transition line. First, there is a transition between a coherent regime (either methylated or demethylated), and a disordered, or mixed, regime (where marked and unmarked nucleosomes coexist with no domain formation). Second, in the coherent regime there is a transition between a methylated phase, when long-range infection is efficient (small $\sigma$), and a demethylated phase when infection is shorter-range (large $\sigma$). Increasing $q_0$ favours the demethylated phase in our parameter range: we interpret this as due to the boundary erasure term which tips the balance in the favour of demethylation, when $\sigma$ is sufficiently large, and for realistic system size, say, of a hundred units/histones.

We have shown that the transition between coherent and mixed regimes can be understood on the basis of a simple stochastic differential equation which is analytically tractable, and which predicts there is a qualitative change in the nature of the effective potential governing the steady state behaviour of the system. At low $q_0$, the potential has two local minima ($m=0$, fully demethylated, or $m=1$, fully methylated), whereas at larger $q_0$ noise dominates and there is a single minimum at intermediate $m$. The transition between methylated and demethylated phase, on the other hand, can be understood qualitatively as a transition between absorbing states in the limit where $q_0\to 0$. In this limit, we have shown that the model is mappable onto a special case of the contact process, known as long-range compact directed percolation, which encompasses the voter model with a long-range interaction. If we modify rates such that only one absorbing states remain, the universality class changes and becomes, as expected, that of ``standard'' (i.e., non-compact) long-range directed percolation, which has been well studied in the literature on contact processes~\cite{Howard}. 

In this $q_0\to 0$ limit, the transition between the two absorbing states (here methylated and demethylated) is a sharp, first-order-like transition, and this appears to be the case also for the transition at $q_0\ne 0$, in the epigenetically coherent phase. The first-order nature of the transition endows the system with bistability and hysteresis close to the critical line~\cite{Erdel2016}. This provides a pathway to the establishment of epigenetic memory -- the phenomenon through which a chromatin region ``remembers'' its state even following a relatively strong perturbation~\cite{MichielettoPRX}. This is different from other $1$D models of epigenetic dynamics~\cite{Sneppen}, where bistability arises due to symmetry breaking in the epigenetically coherent phase, which is either fully methylated or fully acetylated. Interestingly, and in stark contrast with other models, in the unstable regime also the unmarked phase can retain memory of its state. 

It is tempting to speculate that cells may tune the $\sigma$ and $q_0$ parameters, which are associated with the conformational changes of chromatin and the bare affinity of the writing/erasing enzymes, so as to control the variability in the epigentics patterns and thus, in turn, the variability in gene expression within the same population. Recent experiments in yeast support the idea that this may be a biological bet-hedging strategy for survival against random attacks~\cite{Allshire2}. 

\paragraph{Acknowledgements} We thank the European Research Council (ERC CoG Grant
FQ 623 No. 648050 THREEDCELLPHYSICS) for funding.

\section{Appendix A: EFFECTIVE ACTION FOR LRCDP AND POWER COUNTING}
In this Appendix we briefly review the construction of the Reggeon effective field action, specifically for the LRCDP process, through which is possible to individuate an upper critical dimension to our model, when in the presence of absorbing states. Similar actions have been introduced and discussed in previous papers and reviews on both short-range and long-range infection processes~\cite{Vernon,Vernon2,Janssen2,Peliti,Tauber}. If we consider the phenomenological equation in \eqref{LRCDP}, we can introduce the effective field lagrangian density:

\begin{widetext}
\begin{equation}
		\mathcal{L}[\hat{m},m] = \hat{m}(x,t) \left[ (\partial_t - D_A\nabla^\sigma) m(x,t) - \kappa m(x,t)(1-m(x,t)) - \sqrt{\Gamma m(x,t)(1-m(x,t))} \zeta(x,t) \right]
\end{equation}
\end{widetext}
where $ \hat{m} $ is the \textit{response field} and we have omitted any possible short range contribution which would be represented by $ \nabla^2 $ (this term is negligible if $ \sigma < 2 $). The associated effective field action reads:

\begin{equation}
    \mathcal{S} = \int \mathrm{d}t \mathrm{d}^d x \: \mathcal{L}[\hat{m},m].
\end{equation}
By integrating $\mathrm{exp}(-S)$ over the functional measure $\mathcal{D}\hat{m} \mathcal{D}m \mathcal{D}\zeta P\left[\zeta\right]$, one gets the partition function $ \mathcal{Z} $ in the field-theory treatment. The noise $ \zeta $ is gaussianly distributed:

\begin{equation}
	 P \left[\zeta\right] = \mathrm{exp} \left[-\int \mathrm{d} t \mathrm{d}^d x \: \frac{\zeta^2(x,t)}{2} \right].
\end{equation}
The integration over $\mathcal{D}\zeta$ can be explicitly performed, leading to the following effective action:

\begin{widetext}
\begin{equation}
    \mathcal{S} = \int \mathrm{d}t \mathrm{d}^d x \: \hat{m}(x,t) \left[ (\partial_t - D_A\nabla^\sigma) m(x,t) - \kappa m(x,t)(1-m(x,t)) - \frac{\Gamma}{2} \hat{m}(x,t)m(x,t)(1-m(x,t)) \right]
	\label{actionLRCDP}
\end{equation}
\end{widetext}
It is easy to find by a simple power counting the \textit{scaling dimensions} of the coupling constant $ \kappa $ and $ \Gamma $ \cite{Janssen3}. Introducing the length scale $ l^{-1} $, and assuming that $ D_A $ has null naive dimension, we derive:

\begin{equation}
	\begin{split}
	x \sim l^{-1} \mathrm{,} \qquad t^{-1} \sim l^{\sigma} \\
	\hat{m} \sim l^{d} \mathrm{,} \qquad m \sim l^{0} \\
	\kappa \sim l^{\sigma} \mathrm{,} \qquad \Gamma \sim l^{\sigma-d}	
	\end{split}
\end{equation}
For $ \kappa = 0 $ the only coupling constant $ \Gamma $ has null naive dimension if $ d = \sigma $, which demonstrate that the upper critical dimension is $ d_c = \sigma $. As $ d = 1 $ in our model, we expect a mean--field behaviour for $ \sigma < 1 $ (since $ \Gamma $ becomes irrelevant), and a non-trivial behaviour for $ \sigma > 1 $. Note also that the action in Eq.~\eqref{actionLRCDP} is not symmetric under the so-called rapidity-reversal transformation $ \hat{m}(x,t)~\rightarrow~-m(x,-t) $. Instead, for $ \kappa = 0 $, it is symmetric under the transformation $ m(x,t)~\rightarrow 1-m(x,t) $ and $ \hat{m} \rightarrow -\hat{m} $. Clearly, the different symmetry involved produce a different universality class compared with LRDP (for each $\sigma$). 

\section{Appendix B: LRDP effective action and power counting}

In this section we show how our model crosses over the LRDP non-universality class in the limit $ q_\lambda \rightarrow 0$ with $ \bar{\lambda} $ finite. The introduction of a global deletion rate $ q_\mu $ changes the structure of the absorbing states, see the main text. Therefore, the noise in the phenomenological equation becomes $\sqrt{\Gamma m}\zeta$ and the Reggeon effective field action in the Eq.~\eqref{actionLRCDP} can be modified as follows:

\begin{widetext}
\begin{equation}
    \mathcal{S} = \int \mathrm{d}t \mathrm{d}^d x \: \hat{m}(x,t) \left[ (\partial_t - D_A\nabla^\sigma) m(x,t) - \kappa m(x,t)(1-m(x,t)) - q_\mu m(x,t) - \frac{\Gamma}{2} \hat{m}(x,t)m(x,t) \right]
	\label{actionLRDP}
\end{equation}
\end{widetext}

By using the substitution $ \kappa - q_\mu \rightarrow \kappa' $ and $ \kappa \rightarrow g' $, and omitting the primes, we find the effective field action of Eq.~\eqref{LRDP}. Then, by rescaling the field with the length scale $l'$ such that $ \hat{m} \rightarrow l'\hat{m}$ and $ m \rightarrow l'^{-1}m $ with $ l' = \sqrt{2g/\Gamma}$, we obtain 

\begin{widetext}
\begin{equation}
    \mathcal{S} = \int \mathrm{d}t \mathrm{d}^d x \: \hat{m}(x,t) \left[ (\partial_t - D_A\nabla^\sigma) m(x,t) - \kappa m(x,t) - \sqrt{\frac{\Gamma\kappa}{2}} (\hat{m}(x,t)-m(x,t))m(x,t). \right]
	\label{mod_actionLRDP}
\end{equation}
\end{widetext}

Note that the action written in Eq.~\eqref{mod_actionLRDP} satisfies the rapidity-reversal symmetry. By a power counting similar to the one performed in Appendix A, we find

\begin{equation}
	\begin{split}
	x \sim l^{-1} \mathrm{,} \qquad t^{-1} \sim l^{\sigma} \\
	\hat{m} \sim l^{d/2} \mathrm{,} \qquad m \sim l^{d/2} \\
	\kappa \sim l^{\sigma} \mathrm{,} \qquad \Gamma \sim l^{\frac{2\sigma-d}{2}}.
	\end{split}
\end{equation}
Therefore, the upper critical dimension for LRDP is $ d_c = 2\sigma $ \cite{Howard}.

\bibliographystyle{apsrev4-1}
\bibliography{library}

\end{document}